# Generalized Langevin equations and fluctuation-dissipation theorem for particle-bath systems in electric and magnetic fields

Vladimír Lisý and Jana Tóthová

Department of Physics, Faculty of Electrical Engineering and Informatics, Technical University of Košice, Park Komenského 2, 042 00 Košice, Slovakia

ABSTRACT

   The Brownian motion of a particle immersed in a bath of charged particles is considered when the system is placed in magnetic or electric fields. Coming from the Zwanzig-Caldeira-Legget particle-bath model, we modify it so that not only the charged Brownian particle (BP) but also the bath particles respond to the external fields. For stationary systems the generalized Langevin equations are derived. Arbitrarily time-dependent electric fields do not affect the memory functions, the thermal noise force, and the BP velocity correlation functions. In the case of a constant magnetic field two equations with different memory functions are obtained for the BP motion in the plane perpendicular to the field. As distinct from the previous theories, the random thermal force depends on the field magnitude. Its time correlation function is connected with one of the found memory functions through the familiar second fluctuation-dissipation theorem.

**Introduction**

   The Brownian motion (BM) of a tagged particle in a bath of other particles is effectively described by the generalized Langevin equation (GLE) [1]. As distinct from the standard Langevin equation [2], the GLE instead of the Stokes friction force contains a convolution of a memory kernel with the velocity of the tagged particle. The kernel is connected to the random thermal force through the second fluctuation-dissipation theorem (FDT) [1]. Following Kubo [1], in the linear approximation it is usually assumed that if external forces act on the system, they do not affect the thermal force and thus the FDT. The action of such forces is restricted to the Brownian particle (BP), leaving the bath particles unaffected by the external field. However, according to several recent papers such approach is not realistic [3,4,5]. There are a number of important physical problems, where not only the BP but also the particles that constitute the heat bath are subjected to the external field. It has been shown in Ref. [3], by computer simulations, that if the dynamics of a single methane molecule solvated in water is modeled by the GLE, the memory function necessarily depends on the elasticity constant of the external harmonic potential. In Ref. [4], devoted to the study of the influence of external oscillating electric fields on the particle-bath systems, a new form of the FDT was derived: the time correlation of the stochastic force is not just equal to the memory function, but there is an additional term which is proportional to the amplitude of the ac field squared. This term is absent in all previously studied versions of the FDT. In Ref. [5] we considered a stationary particle-bath system in an external harmonic potential within the Zwanzig-Caldeira-Legget (ZCL) model [6,7], derived the GLE and showed that the FDT has the same form as found by Kubo [1]. However, the memory function and the thermal force depend on the elastic constant of the confinement potential.

In the first part of the present short note we revisit the model [4] considering the motion of a charged BP in a bath of charged particles. The system is placed in an electric field which can be constant or arbitrarily dependent on time. It is assumed that not only the BP but also the bath particles respond to the external field. It is shown that the electric force, which does not depend on the positions or velocities of the bath particles, has no effect on the properties of the random thermal force and the memory function. The Kubo's FDT and the BP velocity correlation function thus remain the same as in the absence of the external field. The mean velocity of the tagged BP differs from that in the case when the bath does not respond to the electric force. Within the modified ZCL model, an exact equation for this velocity and its approximate solutions are obtained depending on the frequency distribution of the bath particles regarded as harmonic oscillators. In the second part of the work we consider the same system under a constant magnetic field. As distinct from the electric force, the magnetic force on the charged particles depends on their velocities, which are stochastic variables. We show that the motion of the BP across the field is described by two equations of the GLE type with two memory functions. When the response of the bath to the external field is "switched off", one of these functions disappears. The second one enters the FDT that has the familiar Kubo's form but, if the bath response is taken into account, it depends on the magnetic field magnitude. The obtained explicit expressions for the memory functions are determined by the frequency distribution function of the bath oscillators and their coefficients of coupling to the BP. A method of approximate calculation of the relevant time correlation functions, such as the velocity correlation function (VCF), for the BP in the plane perpendicular to the field is proposed.

**Particle-bath systems in the presence of external electric fields**

Let us first consider the particle-bath system in an external electric field. The particle is linearly coupled with $N$ oscillators, which are not coupled to each other. The electric field with intensity $E(t)$ is oriented along the axis $z$. The bath oscillators with eigenfrequencies $\omega_i$ and masses $m_i$ carry net charges $q_i$, and the charge of the BP of mass $m$ is $Q$. The strength of coupling between the tagged particle and the $i$th oscillator is $c_i$. The momenta of the bath particles and the BP in the $z$ direction are, respectively, $p_i$ and $p$. The Hamiltonian of the system [4,6,7],

$$H = \frac{p^2}{2m} + \frac{1}{2}\sum_{i=1}^{N}\left[\frac{p_i^2}{m_i} + m_i\omega_i^2\left(z_i - \frac{c_i z}{\omega_i^2}\right)^2\right] - E(t)\left(Qz + \sum_{i=1}^{N}q_i z_i\right), \tag{1}$$

consists of the Hamiltonian of the tagged BP in the absence of the external field, the Hamiltonian of the bath of harmonic oscillators coupled to the BP, and the one representing the influence of the electric field on the system. The equations of motion along the direction of the electric field for such a system of particles are

$$\dot{z} = \frac{p}{m}, \quad \dot{p} = \sum_i m_i c_i \left( z_i - \frac{c_i}{\omega_i^2} z \right) + QE(t), \tag{2}$$

$$\dot{z}_i = \frac{p_i}{m_i}, \quad \dot{p}_i = -m_i \omega_i^2 z_i + m_i c_i z + q_i E(t). \tag{3}$$

The terms connected with an external harmonic potential [4-7] can be easily included in the consideration but are not principal for the aims of this work. The equation of motion for the BP can be obtained by substituting the solution of Eqs. (3), $\ddot{z}_i = \dot{p}_i / m_i$, in $\dot{p}$ from (2). This can be effectively done by rewriting the time-dependent quantities in the Laplace transform (LT), e.g., $\tilde{z}(s) = \int_0^\infty dt e^{-st} z(t) = \mathcal{L}\{z(t)\}$. This gives

$$\mathcal{L}\{\dot{p}\} = \sum_i \frac{m_i c_i}{s^2 + \omega_i^2} [sz_i(0) + \dot{z}_i(0)] - \sum_i \frac{m_i c_i^2 s^2}{\omega_i^2 (\omega_i^2 + s^2)} \tilde{z}(s) + \left( Q + \sum_i \frac{q_i c_i}{\omega_i^2 + s^2} \right) \tilde{E}(s). \tag{4}$$

Going back to the time domain with the use of the convolution theorem and the relation $s\tilde{z}(s) - z(0) = \tilde{\upsilon}(s)$ [8], we get the GLE for the velocity of the BP $\upsilon = p/m$,

$$m\dot{\upsilon}(t) + \int_0^t \Gamma(t - \tau)\upsilon(\tau)d\tau = \sum_i \frac{q_i c_i}{\omega_i} \int_0^t \sin[\omega_i(t - \tau)]E(\tau)d\tau + QE(t) + f(t), \tag{5}$$

where $f(t) = \sum_i m_i c_i \left\{ [z_i(0) - c_i \omega_i^{-2} z(0)] \cos(\omega_i t) + \omega_i^{-1} \sin(\omega_i t) \dot{z}_i(0) \right\}$ is the thermal force determined by the random initial quantities $z(0)$, $z_i(0)$, and $\dot{z}_i(0)$ with zero means and $\Gamma(t) = \sum_i m_i c_i^2 \omega_i^{-2} \cos(\omega_i t)$ is the memory function. For the special case $E(t) = E_0 \sin(\omega t)$ Eq. (5) coincides with Eq. (6) from [4] where, however, the first term on the right-hand side of (5) is included in $f(t)$, for which there is no reason. This term does not contain any stochastic quantity and determines an additional effect of the electric field on the BP due to its coupling to the surrounding particles. Neither the memory function $\Gamma(t)$ nor the stochastic force $f(t)$ are affected by the electric force, which is independent on the positions and velocities of the bath particles. In such a view, Kubo's second FDT remains unchanged. As it is seen from (5), what will change is the mean velocity, which will be different from zero and dependent on $E(t)$. The equation for the determination of $V(t) = \langle \upsilon(t) \rangle$ from Eq. (5) is

$$\tilde{V}(s)\left[ms+\tilde{\Gamma}(s)\right]=\left(Q+\sum_i \frac{q_i c_i}{\omega_i^2+s^2}\right)\tilde{E}(s). \tag{6}$$

The VCF $C_{\upsilon\upsilon}(t)=\langle \upsilon(t_0)\upsilon(t_0+t)\rangle=\langle \upsilon(0)\upsilon(t)\rangle$ (the system is conditioned to be stationary) remains the same as in the absence of the field. Multiplying Eq. (5) by $\upsilon(0)$ and averaging with the assumptions $\langle \upsilon(0)f(t)\rangle=0$ (causality) and $\langle \upsilon(0)\rangle=0$, one gets

$$m\dot{C}_{\upsilon\upsilon}(t)=-\int_0^t \Gamma(t-\tau)C_{\upsilon\upsilon}(\tau)d\tau. \tag{7}$$

By using the equipartition theorem, $C_{\upsilon\upsilon}(0)=k_B T/m$, the solution of (5), in the LT, is

$$\tilde{C}_{\upsilon\upsilon}(s)=k_B T/[ms+\tilde{\Gamma}(s)]. \tag{8}$$

If Eq. (5) is multiplied by $f(0)=m\dot{\upsilon}(0)$ and averaged with the use of $\langle \dot{\upsilon}(0)\dot{\upsilon}(t)\rangle=-\ddot{C}_{\upsilon\upsilon}(t)$ and $\langle \dot{\upsilon}(0)\upsilon(t)\rangle=-\dot{C}_{\upsilon\upsilon}(t)$, we find for $C_{ff}(t)=\langle f(t)f(0)\rangle$

$$C_{ff}(t)=-m^2 \ddot{C}_{\upsilon\upsilon}(t)-m\int_0^t \Gamma(t-\tau)\dot{C}(\tau)d\tau, \tag{9}$$

or $\tilde{C}_{ff}(s)=-m^2 s[s\tilde{C}_{\upsilon\upsilon}(s)-C_{\upsilon\upsilon}(0)]$ in the LT. Equations (8) and (9) give $\tilde{C}_{ff}(s)=k_B T\tilde{\Gamma}(s)$, which is the second FDT in its familiar form [1],

$$C_{ff}(t)=k_B T\Gamma(t). \tag{10}$$

Note that in Ref. [4] the right-hand side of Eq. (14) for the correlation function of the thermal force contains an additional term $\sim E(t)E(t')$ due to which this force does not obey the claimed in [4] stationarity.

**Brownian motion in a bath responding to a constant magnetic field**

In the case of an external magnetic field, the situation principally changes. Let the system of the BP and bath particles is placed in a constant magnetic field **B** oriented parallel to the axis $z$. If the positions of the bath particles and the BP are $\mathbf{r}_i=(x_i,y_i,z_i)$ and $\mathbf{r}=(x,y,z)$, and momenta $\mathbf{p}_i=(p_{xi},p_{yi},p_{zi})$ and $\mathbf{p}=(p_x,p_y,p_z)$, respectively, instead of Eqs. (2) and (3) we now have to study the coupled equations of motion with magnetic forces

$$\dot{\mathbf{r}} = \frac{\mathbf{p}}{m}, \qquad \dot{\mathbf{p}} = \sum_i m_i c_i \left( \mathbf{r}_i - \frac{c_i}{\omega_i^2} \mathbf{r} \right) + Q\dot{\mathbf{r}} \times \mathbf{B}, \qquad (11)$$

$$\dot{\mathbf{r}}_i = \frac{\mathbf{p}_i}{m_i}, \qquad \dot{\mathbf{p}}_i = -m_i \omega_i^2 \mathbf{r}_i + m_i c_i \mathbf{r} + q_i \dot{\mathbf{r}}_i \times \mathbf{B}. \qquad (12)$$

Along the axis $z$, the motion of the particles is not affected by the magnetic field and will be not considered here. To get the equations of motion in the plane perpendicular to the field one can act similarly as in obtaining Eq. (4), by using the LT of these equations. So, if $\tilde{y}_i(s)$ obtained from Eq. (12) is substituted into (11), one gets for $\tilde{x}_i(s)$ the equation

$$\left( s^2 + \omega_i^2 + \frac{s^2 \Omega_i^2}{s^2 + \omega_i^2} \right) \tilde{x}_i(s) = s \left( 1 + \frac{\Omega_i^2}{s^2 + \omega_i^2} \right) x_i(0) + \dot{x}_i(0)$$

$$+ \Omega_i \left( \frac{s^2}{s^2 + \omega_i^2} - 1 \right) y_i(0) + \frac{s\Omega_i}{s^2 + \omega_i^2} \dot{y}_i(0) + c_i \tilde{x}(s) + \frac{sc_i\Omega_i}{s^2 + \omega_i^2} \tilde{y}(s). \qquad (13)$$

A similar equation for $\tilde{y}_i(s)$ is obtained if all $x$ are replaced by $y$ and vice versa, and the cyclotron frequencies $\Omega_i = q_i B / m_i$ are changed to $-\Omega_i$. The equations contain initial positions of the oscillators, $x_i(0)$, $y_i(0)$, and their initial velocities, $\dot{x}_i(0)$, $\dot{y}_i(0)$, all of which are random quantities with zero mean. Note that in the absence of the external field the inverse LT [8] immediately gives the known solution of the inhomogeneous equation (12) [6]

$$x_i(t) = x_i(0) \cos(\omega_i t) + \frac{\dot{x}_i(0)}{\omega_i} \sin(\omega_i t) + \frac{c_i}{\omega_i} \int_0^t x(t') \sin[\omega_i(t-t')] dt'. \qquad (14)$$

Continuing the calculations in the LT, we substitute the solutions for $\tilde{x}_i(s)$ and $\tilde{y}_i(s)$ in the Laplace transformed Eq. (11) and only after that we return to the time domain. The final equations for the BP are

$$m\dot{\upsilon}_x(t) = QB\upsilon_y(t) - \int_0^t \upsilon_x(t') G(t-t') dt' + \int_0^t \upsilon_y(t') H(t-t') dt' + f_x(t), \qquad (15)$$

$$m\dot{\upsilon}_y(t) = -QB\upsilon_x(t) - \int_0^t \upsilon_y(t') G(t-t') dt' - \int_0^t \upsilon_x(t') H(t-t') dt' + f_y(t). \qquad (16)$$

Here, $f_x(t)$ and $f_y(t)$ are independent zero-mean projections of the random force that are determined by the initial positions and velocities of the particles in the system. It is seen that (15) and (16) are of the GLE type but with two functions, $G(t)$ and $H(t)$, determining the retardation effects in the BP dynamics. If we denote $\gamma_i = (\Omega_i^2 + 4\omega_i^2)^{1/2}$ and $\gamma_i^\pm = \gamma_i \pm \Omega_i$, these functions are expressed as

$$G(t) = 2\sum_{i=1}^{N} \frac{m_i c_i^2}{\gamma_i} \left[ \frac{\cos(\gamma_i^+ t/2)}{\gamma_i^+} + \frac{\cos(\gamma_i^- t/2)}{\gamma_i^-} \right], \tag{17}$$

$$H(t) = 2\sum_{i=1}^{N} \frac{m_i c_i^2}{\gamma_i} \left[ \frac{\sin(\gamma_i^+ t/2)}{\gamma_i^+} - \frac{\sin(\gamma_i^- t/2)}{\gamma_i^-} \right]. \tag{18}$$

Equations (15) and (16) possess large possibilities for the predictions of the behavior of systems of charged particles in the magnetic field. Specific solutions for the correlation functions describing the random motion of the BPs such as the VCF $C_{\upsilon\upsilon}(t) = \langle \upsilon(t_0)\upsilon(t_0+t) \rangle$) require knowledge of the distribution function for the frequencies $\{\omega_i\}$ and coupling constants $c_i$ for concrete systems. However, some results of general character can be obtained just assuming that the system is conditioned to be stationary, without restricting it to be in thermal equilibrium [9].

So, let us multiply Eqs. (15) and (16) by $\upsilon_x(0)$ and $\upsilon_y(0)$, respectively, and statistically average. From the obtained equations for $C_{\upsilon_\alpha \upsilon_\alpha}(t)$ ($\alpha \equiv x, y$), $C_{\upsilon_x \upsilon_y}(t) = \langle \upsilon_x(t)\upsilon_y(0) \rangle$ and $C_{\upsilon_y \upsilon_x}(t) = \langle \upsilon_y(t)\upsilon_x(0) \rangle$ with $C_{\upsilon_x \upsilon_y}(0) = 0$, we again turn to the LT and, with the commonly used causality principle due to which $\langle \upsilon_\alpha(0) f_\alpha(t) \rangle = 0$ for $t > 0$, applying the convolution theorem [8] and equipartition, $C_{\upsilon_\alpha \upsilon_\alpha}(0) = k_B T / m$, we find the VCFs

$$\tilde{C}_{\upsilon_\alpha \upsilon_\alpha}(s) = k_B T \frac{ms + \tilde{G}(s)}{[ms + \tilde{G}(s)]^2 + [QB + \tilde{H}(s)]^2}, \tag{19}$$

$$\tilde{C}_{\upsilon_x \upsilon_y}(s) = -\tilde{C}_{\upsilon_y \upsilon_x}(s) = k_B T \frac{QB + \tilde{H}(s)}{[ms + \tilde{G}(s)]^2 + [QB + \tilde{H}(s)]^2}. \tag{20}$$

These functions are related to all other relevant correlation functions, such as the positional autocorrelation function in the $x$ direction in the plane perpendicular to the field, $C_{xx}(t) = \langle x(0)x(t) \rangle$, the mean square displacement $X(t) = 2[C_{xx}(0) - C_{xx}(t)]$ or the time-dependent diffusion coefficient $D_x(t) = \dot{X}(t)/2$. In the LT, $\tilde{X}(s) = 2\tilde{D}_x(s)/s = 2\tilde{C}_{\upsilon_x \upsilon_x}(s)/s^2$ [10]. Analogous relations hold for the $y$ direction. Moreover, if we multiply Eq. (15) by $f_x(0) = m\dot{\upsilon}_x(0) - QB\upsilon_y(0)$ and (16) by $f_y(0) = m\dot{\upsilon}_y(0) + QB\upsilon_x(0)$ and average the results, the time correlation functions of the random forces $C_{f_\alpha f_\alpha}(t) = \langle f_\alpha(t) f_\alpha(0) \rangle$ can be expressed through the already derived VCFs. Due to stationarity, the following identities can be used for $C_{\upsilon\upsilon}(t)$ ($\upsilon$ stays for $x$ and $y$ components of the velocity): $\langle \dot{\upsilon}(t)\dot{\upsilon}(0) \rangle = -\ddot{C}_{\upsilon\upsilon}(t)$, $\langle \upsilon(t)\dot{\upsilon}(0) \rangle = -\dot{C}_{\upsilon\upsilon}(t)$,

$\langle \dot{v}(t)v(0)\rangle = \dot{C}_{vv}(t)$, $\dot{C}_{vv}(0) = 0$. For the different projections of the velocity it holds $\langle v_y(t)\dot{v}_x(0)\rangle = -\langle \dot{v}_y(t)v_x(0)\rangle = -\dot{C}_{v_y v_x}(t)$, $C_{v_x v_y}(0) = 0$. By using these identities, we obtain

$$\tilde{C}_{f_\alpha f_\alpha}(s) = k_B T \tilde{G}(s), \ \alpha = x, y. \tag{21}$$

The second FDT in its usual form thus again holds also in the case when the bath responds to the external magnetic field. The important difference from Kubo's relation is in the explicit dependence of the memory function on the external field. Here the role of the memory function is played only by one of the functions (17) and (18). Note that when there is no response of the bath, $\Omega_i = 0$, $H(t) = 0$, and Eqs. (15-21) coincide with the previously used equations for the BM of particles in a magnetic field with the memory function $G(t)$ [11,12].

**Conclusion**

The second FDT was derived by Kubo for the GLE assuming that random thermal force (thermal noise) is not affected by the external harmonic field. The elastic force was skipped from the equation. In Ref. [5] we have shown that for stationary systems the Kubo's FDT is also valid for the full GLE with the external elastic force and that the thermal force and the memory function may be affected by it. In the present work, we derived the GLEs coming from the ZCL model modified to the cases when the bath responses to external (time-dependent) electric or (constant) magnetic fields. In both the situations the familiar form of the FDT remains unchanged, but while the electric forces have no effect on the memory function, the magnetic field affects it as well as the thermal force. The found equations of motion for the BP allow obtaining analytical expressions for the memory functions and relevant time correlation functions (for some attempts in this direction see the recent papers [13-15]). The presented model possesses various possibilities for its further development and designing and interpretation of new experiments. The next steps towards a more realistic description of the Brownian motion under external forces should be the inclusion into the consideration nonlinear effects due to the interactions between the bath particles, which are neglected in the presented calculations. The nonlinear character of the stochastic environment of the (harmonically trapped) BP has already been effectively taken into account in the recent work [16] but with no response of the bath to the external force.

**Acknowledgment**
This work was supported by the Scientific Grant Agency of the Slovak Republic through grant VEGA 1/0250/18.

**References**

[1] Kubo R, Rep Progr Phys 1966;29:255. https://doi.org/10.1088/0034-4885/29/1/306

[2] Langevin P, C R Acad Sci (Paris) 1908;146:530. English translation: Lemons DS, Amer J Phys 1997;65:1079. https://doi.org/10.1119/1.18725

[3] Daldrop JO, Kowalik BG, Netz RG, Phys Rev X 2017;7:041065. https://doi.org/10.1103/PhysRevX.7.041065

[4] Cui B, Zaccone A, Phys Rev E 2018;97:060102(R). https://doi.org/10.1103/PhysRevE.97.060102

[5] Lisý V, Tóthová J, Results Phys 2019;12:1212. https://doi.org/10.1016/j.rinp.2019.01.003

[6] Zwanzig R, Nonequilibrium Statistical Mechanics. Oxford University Press, New York; 2001. https://scholar.google.com/scholar?cluster=12121170298723806 79&hl=sk&as_sdt=0,5

[7] Caldeira AO, Legget AJ, Ann Phys 1983;149:374. https://doi.org/10.1016/0003-4916(83)90202-6

[8] Abramowitz A, Stegun IA, Handbook of Mathematical Functions. National Bureau of Standards, Washington, DC; 1964. http://people.math.sfu.ca/~cbm/aands/frameindex.htm

[9] Donado F, Moctezuma RE, López-Flores L, Medina-Noyola M, Arauz-Lara JL, Sci Rep 2017;**7**:12614. https://doi.org/10.1038/s41598-017-12737-1

[10] Tóthová J, Vasziová G, Glod L, Lisý V, Eur J Phys 2011;32:645. https://doi.org/10.1088/0143-0807/32/3/002

[11] Paraan FNC, Solon MP, Esguerra JP, Phys Rev E 2008;77: 022101. https://doi.org/10.1103/PhysRevE.77.022101

[12] Lisý V, Tóthová J, Transport Theory Stat Phys 2014;42:365. https://doi.org/10.1080/00411450.2014.922480

[13] Tóthová J, Šoltýs A, Lisý V, J Mol Liq 2020; 317: 113920. https://doi.org/10.1016/j.molliq.2020.113920

[14] Lisý V, Tóthová J, Acta Phys Pol A 2020;137:657. https://doi.org/10.12693/APhysPolA.137.657

[15] Tóthová J, Lisý V, Physica A 2020;559:125110. https://doi.org/10.1016/j.physa.2020.125110

[16] Müller B, Berner J, Bechinger C, Krüger M, New J Phys 2020; 22:023014. https://doi.org/10.1088/1367-2630/ab6a39